\documentclass{article}
\usepackage{spconf,amsmath,graphicx}
\usepackage{amssymb,mathtools}
\usepackage{steinmetz} 
\usepackage{amsthm}
\usepackage{cite}
\usepackage{url} 
\usepackage{color} 
\usepackage{subcaption}
\usepackage{gensymb}
\usepackage{enumitem}

\title{Smart CSI Processing For Accurate Commodity WiFi-based Humidity Sensing}
\name{Yirui Deng, \,Deepak Mishra, \,Shaghik Atakaramians, \,and\, Aruna Seneviratne\thanks{This work is supported by the Australian Research Council Discovery Early Career Researcher Award (DE230101391) and Australian National Intelligence and Security Discovery Research Grants (NI230100096).}}
\address{School of Electrical Engineering and Telecommunications, University of New South Wales, Australia\\
Emails: \{yirui.deng, d.mishra, s.atakaramians, a.seneviratne\} @unsw.edu.au}
\begin{document}
\ninept
\maketitle
\begin{abstract}
\vspace{-1mm}
Indoor humidity is a crucial factor affecting people's health and well-being. Wireless humidity sensing techniques are scalable and low-cost, making them a promising solution for measuring humidity in indoor environments without requiring additional devices. Such, machine learning (ML) assisted WiFi sensing is being envisioned as the key enabler for integrated sensing and communication (ISAC). However, the current WiFi-based sensing systems, such as WiHumidity, suffer from low accuracy. We propose an enhanced WiFi-based humidity detection framework to address this issue that utilizes innovative filtering and data processing techniques to exploit humidity-specific channel state information (CSI) signatures during RF sensing. These signals are then fed into ML algorithms for detecting different humidity levels. Specifically, our improved de-noising solution for the CSI captured by commodity hardware for WiFi sensing, combined with the k-th nearest neighbour ML algorithm and resolution tuning technique, helps improve humidity sensing accuracy. Our commercially available hardware-based experiments provide insights into achievable sensing resolution. Our empirical investigation shows that our enhanced framework can improve the accuracy of humidity sensing to $97\%$.
\end{abstract}
\vspace{-1mm}
\begin{keywords}
Integrated sensing and communication, machine learning, WiFi-based sensing, humidity sensing 
\end{keywords}
\vspace{-1mm}
\vspace{-1mm}
\section{Introduction}
\label{sec:intro}
\vspace{-2mm}
The management of indoor humidity levels plays an important role in occupant health and well-being. As suggested by the National Asthma Council Australia, the recommended indoor humidity is $30\%$ to $50\%$ to avoid mould, dust mites, and vulnerability to airborne viruses like the flu \cite{humidlevel}. The rapid deployment of Internet-of-things (IoT) technologies enables the continuous measurement and autonomous control of indoor humidity levels with simple and low-cost devices. On the other hand, recent research on commodity WiFi-based sensing techniques \cite{commodityWiFi,ACMCovid} demonstrated its capability as a device-free alternative for sensing environment parameters such as temperature \cite{Li2021}, wetness~\cite{yiruiwetness}, and its advantages of infrastructural ubiquity. Thus, there is a huge potential for WiFi-based sensing techniques in humidity-sensing applications. Specifically, with WiFi sensing-based humidity detection technology, the goal of integrated sensing and communications (ISAC) in the sixth generation (6G) of wireless networks can be realised where ambient WiFi signals can also be used for indoor environment monitoring~\cite{9737357}. 
So, it is necessary to examine the effectiveness of such Radio Frequency (RF) sensing techniques in terms of resolution for various ISAC applications. 
\vspace{-1mm}
\subsection{State-of-the-Art}
\label{subsec:sota}
\vspace{-1mm}
Modern humidity sensors, or hygrometers, correlate the humidity level with the variation of electrical properties, such as capacitance or resistance. Such techniques require dedicated devices and suffer from incurred costs like deployment overhead, power consumption and maintenance. Recent research on Integrated Sensing and Communication (ISAC) aims to address this issue by using the attenuati-

\noindent on of a radio-frequency (RF) wave to estimate the humidity level along its propagation paths. Various prior studies like \cite{Dai2018mmhumidity} and \cite{wang2023} demonstrated the feasibility of humidity sensing in indoor environments with the channel state information (CSI) of RF signals and machine learning (ML) techniques, and \cite{mejia2023hyperlocal} extended its application to outdoor. However, most of the existing RF propagation-based humidity sensing research was based on higher frequency signals such as millimetre wave (mmWave) due to its good sensitivity, in terms of significant, easily-measurable attenuation responses to humidity variations. However, specific to indoor applications, the deployment of such technology is still constrained by the mmWave devices which are not readily available for most households.

In contrast, WiFi infrastructure is ubiquitous in modern society and covers most indoor environments. As suggested by the developing IEEE WiFi standard \cite{restuccia2021ieee}, it is foreseeable that all WiFi devices will be equipped with the capabilities for CSI-based sensing~\cite{en17020485}. Thus, WiFi-based humidity sensing has the advantage in indoor applications of a single sensor and a distributed sensor network that is readily available and cost-free. Existing research for RF-propagation-based humidity sensing with WiFi signal is still limited and primitive. WiHumidity \cite{zhang2017wihumidity} carried out a proof-of-concept study and proposed a framework using WiFi CSI and Support Vector Machine (SVM) classifier to acquire a feasible prediction of humidity levels. However, the average prediction accuracy is $79\%$, which is less than satisfactory for practical deployment.

\vspace{-1mm}
\subsection{Motivation and Contributions}
\label{subsec:moticont}
\vspace{-1mm}
As elaborated above, the WiFi-based humidity sensing technique has great potential in indoor humidity management applications due to its advantages of ubiquitous infrastructure, minimum operating cost and large scalability to distributed humidity sensor networks with the support of future standardized WiFi equipment. However, the existing framework, such as WiHumidity, suffers from low accuracy and to our best knowledge, the enhancement to this framework has not been investigated much. Therefore, it is necessary to enhance the performance of the sensing framework, in terms of prediction accuracy, to a satisfactory level before it can be practically deployed. 

The three-fold contribution of this work is summarized below.
\vspace{-2mm}
\begin{itemize}[itemsep=0pt,parsep=0pt,leftmargin=*]
    \item We provided insight into the workflow pipeline for getting accurate prediction outcomes from raw CSI data collected, including denoising, ML model selection and resolution tuning. We proposed a novel de-noising approach and presented optimal ML algorithm selection and resolution tuning to significantly increase underlying prediction accuracy.
    \item We demonstrated the applicability and reproducibility of the proposed accuracy improvement technique using commercial off-the-shelf hardware for experimental validation. This empirical verification also speeds up the adoption of this sensing technology. 
    \item We provided valuable design insights into the impact of humidity sensing resolution on the prediction accuracy of our technology.
\end{itemize}

\vspace{-3mm}
\section{Experiment Setup Details}
\label{sec:setup}
\begin{figure}[!t]
    \centering
    \includegraphics[width=3.4in]{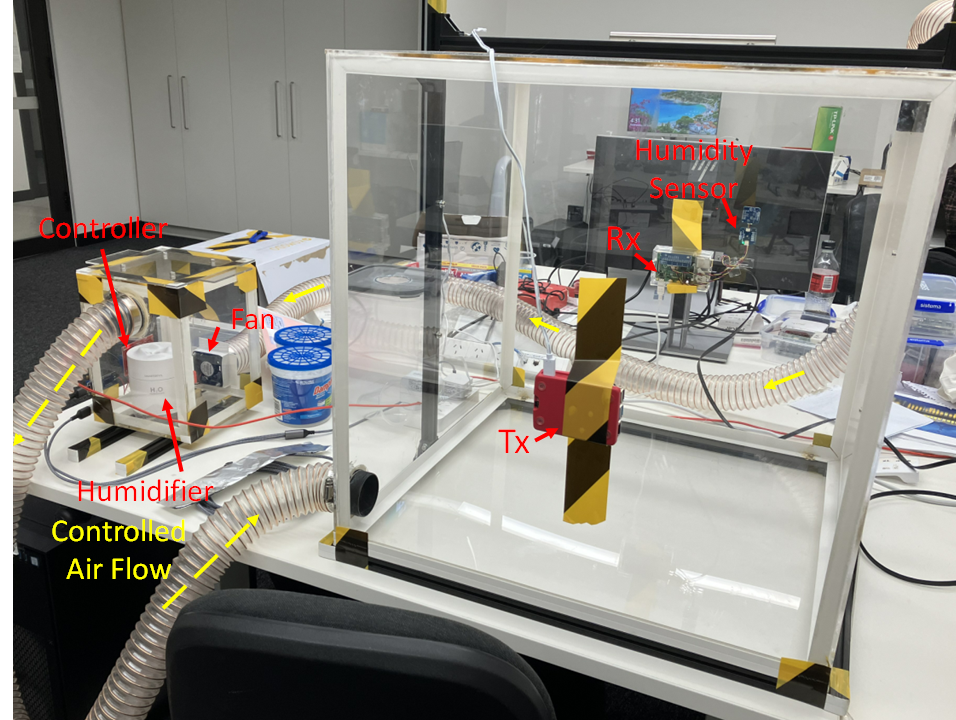}\vspace{-2mm}
    \caption{Experiment Setup.}\vspace{-2mm}
    \label{fig:setup}
\end{figure}

\vspace{-1mm}
We start by presenting the details of the experimental setup for the WiFi-based humidity sensing system. To monitor the WiFi CSI characteristics under an environment with controlled humidity variation, We constructed an acrylic enclosed chamber of $600\times600\times600$mm as shown in Fig. \ref{fig:setup}. The humidity manipulation was achieved via a separate container containing a humidifier, and a fan was connected to the chamber with hoses of $60$mm, forming a closed air-circulation system. The humidifier and the fan were controlled by a remote-accessible Arduino controller. A pair of Raspberry Pi 4B (RPi) with a single antenna were mounted on the opposing surfaces of the chamber as the transmitter (Tx) and receiver (Rx) nodes for CSI extraction. The Rx RPi was loaded with a Nexmon \cite{Nexmon} CSI extraction tool, which measured the CSI from received frames sent by Tx with a $5.18$GHz centre frequency and $80$MHz bandwidth. In addition, a BME280 humidity sensor was mounted within the chamber, which recorded real-time relative humidity readings as data labels for ML.

The experiment was conducted in a laboratory room that was air-conditioned and maintained a consistent temperature. Therefore, to simulate different humidity levels, we used the above-mentioned acrylic chamber during the training and prediction phases. However, we can generalize our results to indoor environments that have sufficient WiFi coverage. Multiple trials of the experiment were conducted at different times and days. For each trial, the humidity was controlled to gradually rise by $1\%$ steps from the ambient room humidity of $40\%$. After each step change, the system was stabilized while the humidifier and fan were off, and then CSI was measured and recorded. After the humidity reached about $70\%$, the chamber was set to depletion mode where the humidity inside the chamber took about 8 hours to gradually reduce to ambient level. The transmission rate of Tx was configured as $4$Hz, so the sample rate of Rx was about 8 frames per second, including the PING frames triggered by Tx and ACK frames generated in hand-shaking.

\vspace{-2mm}
\section{Proposed Humidity Sensing Framework}\vspace{-2mm}
\label{sec:framework}
\begin{figure}[!t]
    \centering
    \includegraphics[width=0.4\textwidth]{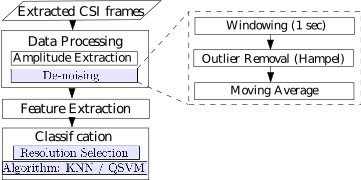}\vspace{-2mm}
    \caption{Workflow of proposed enhanced Humidity Sensing System. }
    \label{fig:workflow}
\end{figure}

The overall workflow of the proposed framework is as shown in Fig. \ref{fig:workflow}. For each frame collected, the CSI amplitude for each subcarrier is extracted from the raw complex value with amplitude and phase information coupled and then de-noised to form the CSI dataset. For a fair comparison, we applied the same standardization and feature extraction mechanism as WiHumidity, in which seven different features such as mean, standard deviation, median absolute deviation, interquartile range, maximum value, skewness and entropy are calculated across subcarriers, and appended to the CSI dataset as the 243rd to 249th feature vectors. It is worth mentioning that since our data was collected for every $1\%$ level of humidity, instead of directly using the exact humidity as the label, we have the freedom to modify the resolution by binning the humidity level via the equation
\begin{equation}
    h_l = n (\text{round}(h/n)),
\end{equation}
where $n$ denotes the size of each bin, also recognized as the resolution, $h$ denotes the exact humidity level, and $h_l$ denotes the bin it is associated to. Then, the different bins can be used as different classes for the classifiers. Lastly, the combined dataset is used to train a classification model with a preset resolution and a specific algorithm.  Although the overall workflow structure seems inherited from WiHumidity, we proposed three novel enhancements, as indicated in the shadowed boxes of Fig. \ref{fig:workflow} to improve the accuracy performance, which will be elaborated on in detail as follows.

\vspace{-2mm}
\subsection{Proposed De-noising Method for Nexmon Extracted CSI}
\label{subsec:denoise}
\begin{figure}[!t]
    \centering
    \includegraphics[width=0.35\textwidth]{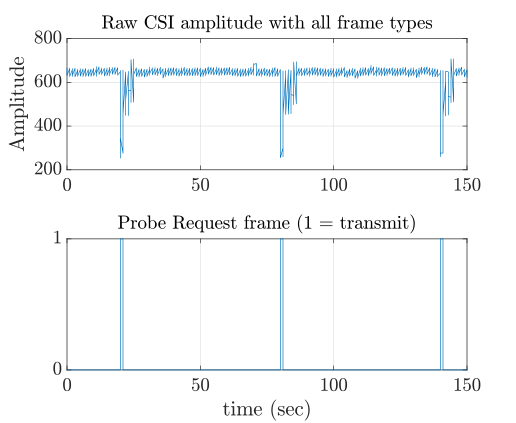}
\vspace{-2mm}
    \caption{CSI disturbance introduced by Probe Request frames. }
    \label{fig:proberequest}
\end{figure}

\vspace{-1mm}
Prior studies like \cite{Dai2018mmhumidity} elaborated that the noise in the received CSI can significantly affect the judgement accuracy of classifiers. The previous WiFi sensing methods did not address the issue of noise, as the variables they were attempting to learn had distinct CSI signatures for different categories, even in the presence of noise. However, for WiFi-based humidity sensing, whose sensitivity to humidity (in terms of having unique CSI signatures for different humidity levels) is relatively lower, de-noising becomes more crucial. As the exact de-noising method applied in WiHumidity was not explicitly mentioned, it is reasonable that a carefully designed de-noising mechanism can potentially improve prediction accuracy. To achieve that, we started with close examinations of the received CSI frames extracted using Nexmon tool \cite{Nexmon} as shown in Fig. \ref{fig:proberequest}. It is observable that there are two types of noise components: a high-frequency noise with small magnitude, and a large-magnitude disturbance that occurs periodically. By checking the individual frames received, we noticed that during the WiFi transmission, Tx sent out a Probe Request (PR) frame every 60 seconds, leading to significant disturbances to the CSI amplitudes of all frames captured by Rx in the subsequent 5 to 10 seconds, as shown in Fig. \ref{fig:proberequest}. While leaving the in-depth investigation for future studies, we considered these affected data as outliers and decided to remove them with a Hampel filter.

\begin{figure}[!t]
    \centering
    \includegraphics[width=0.35\textwidth]{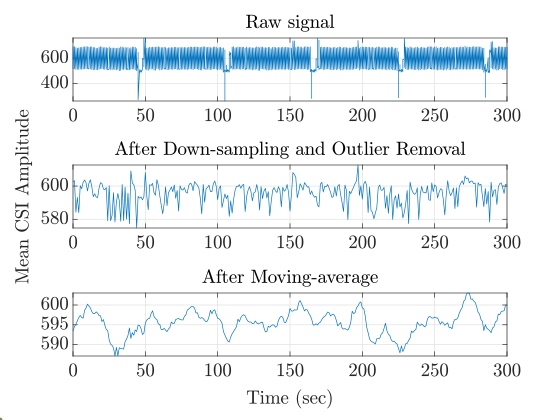}
\vspace{-3mm}
    \caption{Effect of Hampel filters and moving average}
    \label{fig:filter-effect}
\end{figure}
The complete denoise workflow is illustrated as Fig \ref{fig:workflow}. Given the fact that the variation of humidity is gradual, while the WiFi frames are sampled at a faster rate, we performed a down-sampling by windowing-average the received frames before applying the outlier removal. Thus,  without losing information, the outlier removal can be realized with a Hampel filter with a shorter length and the subsequent ML is less computationally heavy. After the data was filtered with the Hampel filter, we further applied a moving average with a window length of 10 to remove the high-frequency noise component. The effect of the de-noising process is demonstrated by a 5-minute example as shown in Fig. \ref{fig:filter-effect}. Three subplots show the mean CSI amplitude across all 242 subcarriers before de-noising, after outlier removal, and after the entire de-noising, respectively. It is observable that not only the periodical large disturbances are removed, but the number of small fluctuations and the overall variance of the mean CSI are reduced. This filtering plays a key role in processing the CSI amplitudes before feeding them to the ML algorithm.

\vspace{-1mm}
\subsection{Selection of ML Algorithm for Humidity Detection}
\label{subsec:selML}
\begin{figure}[!t]
    \centering
    \includegraphics[width=0.35\textwidth]{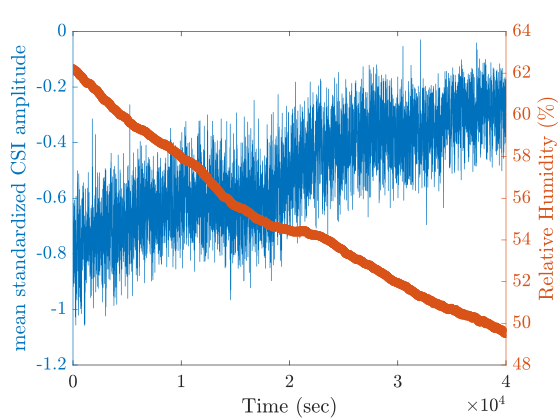}
\vspace{-2mm}
    \caption{An example of the correlation between mean CSI amplitude and the relative humidity from one experiment trial.}
    \label{fig:correlation}
\end{figure}

\vspace{-1mm}
The outcome from one of our experiment trials is presented in Fig. \ref{fig:correlation}, which provides a glance of CSI changes with respect to the humidity change. The blue line represents the mean of standardized CSI amplitude after de-noising across all subcarriers, while the red thick line indicates the respective humidity measured. The method we used for standardization is the z-score that
\begin{equation}
\label{fig:zscore}
\text{z-score(}x\text{)} = \frac{x - E[x]}{\sigma(x)},
\end{equation}
where $E[x]$ denotes the mean or expectation of random quantity $x$ and $\sigma(x)$ denotes the standard deviation of $x$. It is clearly observable that there is a negative correlation between the CSI amplitude and the humidity, and we verified that this observation holds for data from other experiment trials as well. We follow the classification approach for the ML instead of the regression approach because in the practical deployments, the accuracy of the label data is limited by the capability of the device, which could introduce errors to regression. More details will be discussed in the following section.

Although in Fig. \ref{fig:correlation}, the correlation appears approximately linear, the empirical studies from the literature reveal that the correlation is indeed non-linear. mmHumidity \cite{Dai2018mmhumidity} stated that the attenuation of the propagated signal is proportional to the square of humidity and WiHumidity \cite{zhang2017wihumidity} considered the attenuation as a product of two humidity-related components. Thus, we considered exploring the classifiers with non-linear kernels, such as quadratic SVM (QSVM) and k-nearest neighbour (KNN), which are expected to produce higher prediction accuracy compared to the linear SVM (LSVM) used in WiHumidity for this case. We will show via our empirical investigation that this selection of ML design plays a key role in WiFi-based humidity sensing performance.  

\vspace{-1mm}
\subsection{Selection of Humidity Sensing System Resolution}
\vspace{-1mm}
\label{subsec:resolution}
Other than the two enhancement techniques discussed above, we also considered the impact of humidity resolution configuration on the prediction accuracy of such a sensing framework. As elaborated in Section \ref{sec:framework}, the resolution is defined as the humidity difference between the adjacent classes, that is, the bin size while binning the humidity labels into classes.

As the framework relies on a supervised learning process, the correctness of the label is essential for the outcome of the prediction, while the correctness of the humidity labels depends on the precision of the measuring device. For example, the devices we used for measuring the humidity for labelling have a precision of $\pm2\%$ relative humidity as shown on the datasheet \cite{bme280datashet}, which is common to most of the digital sensors on the market. Thus, expecting a resolution higher than $4\%$ can have a negative impact on prediction accuracy due to poor labelling. On the other hand, lower resolution can increase the prediction accuracy since the classifier is no longer striving for finer discrepancies between classes, which might lead to model overfitting. As long as the specification of the applications allows, a reasonable resolution configuration can achieve a better balance of prediction accuracy and system sensitivity.

In short, we proposed three enhancement techniques, including applying a carefully designed de-noising process, using non-linear-kernel classifiers instead of linear approximated LSVM, and choosing a proper resolution of the framework. As WiHumidity did not report details of de-noising and resolution used, we used WiHumidity's reported accuracy as a benchmark and evaluated the outcomes of our proposal via thorough experimental trials with different combinations of criteria. The empirical comparison details will be elaborated on in the next section.

\vspace{-2mm}
\section{Empirical Performance Evaluation}
\vspace{-2mm}
\label{sec:numresult}
\begin{figure}[!t]
    \centering
    \includegraphics[width=0.35\textwidth]{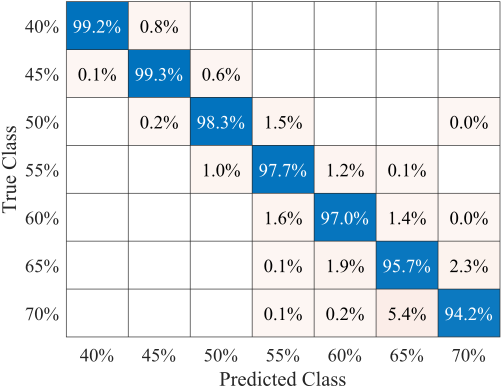}
\vspace{-2mm}
    \caption{An example of confusion matrix from one classification session of 5\% resolution and KNN algorithm.}
    \label{fig:CM_sample}
\end{figure}
With the dataset from multiple experimental trials, we randomly selected 50\% of the data for training the model and the rest 50\% for evaluation. Fig. \ref{fig:CM_sample} illustrated an example of the prediction outcome of humidity from $40\%$ to $70\%$ with KNN and 5\% resolution. It is evident that our proposed techniques have significantly improved the accuracy of humidity prediction when compared to the techniques used in WiHumidity. The enhanced framework is now capable of predicting humidity with an average accuracy of $97.4\%$, which is a $23\%$ increase from WiHumidity's average accuracy of $79\%$. With this level of accuracy, the framework can be deemed appropriate for most monitoring and control applications.

\begin{figure}[!t]
    \centering
    \includegraphics[width=0.4\textwidth]{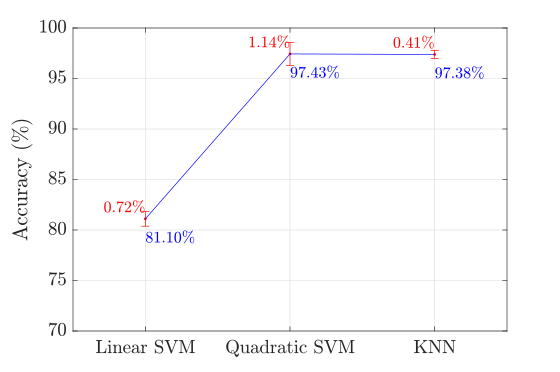}
\vspace{-5mm}
    \caption{Average and variation of prediction accuracy using different classification algorithms with a resolution of 5\%.}
    \label{fig:eval_5pct}
\vspace{-5mm}
\end{figure}
In addition, to evaluate the robustness of the results above, we engaged multiple rounds of data selection and classifications with LSVM, QSVM and KNN under the same condition. Fig. \ref{fig:eval_5pct} shows the performances of each individual algorithm, where the average accuracy values are indicated in blue and the spans of accuracy variation are indicated in red. We can observe that LSVM has an average accuracy of $81.1\%$ which is very close to WiHumidity's performance of $79\%$, but the variation is much smaller as WiHumidity's accuracy was from $69\%$ to $91\%$. It could be due to the properly designed de-noising reduced the randomness of dataset, resulting in less variation due to training data selection. On the other hand, QSVM and KNN outperform in terms of average accuracy. Both non-linear algorithms are able to reach more than $97\%$ accuracy, which validates our elaboration in Section \ref{subsec:selML} that a non-linear classification algorithm can be expected for accuracy improvement. Moreover, the literature revealed that the computational complexity of KNN is $\mathcal{O}(dn)$ \cite{CunninghamKNN} while the naive SVM implementation has a complexity of $\mathcal{O}(n^2)$ \cite{CERVANTES2020189}, where $d$ denotes the number of features and $n$ denotes the number of samples. Thus, KNN is well-suitable as our first option of classifier algorithm for the humidity sensing framework due to potentially less computational complexity.

\begin{figure}[!t]
    \centering\vspace{-5mm}
    \includegraphics[width=0.4\textwidth]{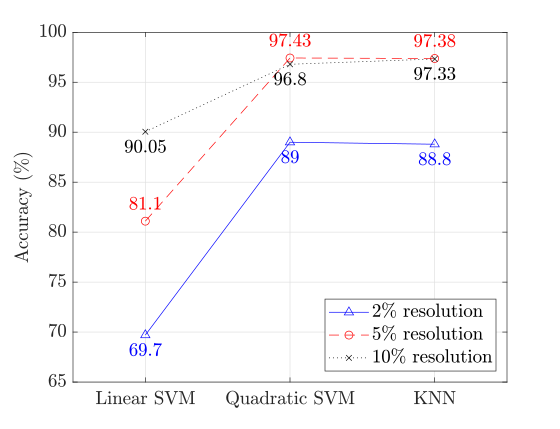}
\vspace{-5mm}
    \caption{Average prediction accuracy using LSVM, QSVM and KNN with different resolutions.}
\vspace{-4mm}
    \label{fig:eval_all}
\end{figure}
Lastly, we evaluated the impact of resolution by classifying the data with $2\%$, $5\%$ and $10\%$ resolution, respectively, as shown in Fig. \ref{fig:eval_all}. It is observable that $2\%$ resolution, which is indicated as a solid blue line, is overly stringent, which resulted in significant accuracy degradation for all algorithms. This validated the statement in Section \ref{subsec:resolution} that non-reliable labelling can reduce the classification performance regardless of the algorithm used since the ground-truth data was collected with $\pm 2$\% accuracy. In contrast, the $5\%$ resolution indicated by the red dashed line, which is close to the resolution of the ground truth, seems with a reasonable performance with our proposed techniques for practical applications. Also, the accuracy of LSVM is improved by further relaxing the resolution to $10\%$, as shown by the black dotted line. This validated the elaboration in Section \ref{subsec:resolution} that lower resolution can improve the accuracy when the classifier is incapable of differentiating the fine discrepancies between classes at higher resolution. In contrast, there are no further accuracy raises for QSVM and KNN from $5\%$ to $10\%$ resolution, as the classification with $5\%$ resolution is within their capability. 


\vspace{-3mm}
\section{Conclusion}
\label{sec:conclusion}
\vspace{-2.5mm}
WiFi-based humidity sensing has enormous potential for indoor humidity management applications due to its advantages of ubiquity, device-free operation, and low cost. However, the existing WiFi-based humidity sensing framework, such as WiHumidity, has less than adequate prediction accuracy for practical deployments. To address this issue, we proposed a new WiFi sensing framework for humidity detection that provided three-fold enhancement in terms of improved de-noising, KNN classifier, and properly tuned resolution. Through experimental validation using commodity hardware, we found that the enhanced framework can effectively achieve a satisfactory accuracy of $97\%$, which is adequate for most indoor applications. In future studies, we plan to expand the coverage of our humidity sensing technology using a network of distributed WiFi devices and investigate the performance of our framework in a more realistic environment having interference from other WiFi devices.

\bibliographystyle{IEEEbib}
\bibliography{refs}

\end{document}